\documentclass[showpacs, oneside, twocolumn, prl, amsmath, amssymb, nofootinbib, superscriptaddress]{revtex4-1}

\usepackage{cases}
\usepackage{amsmath}
\usepackage{amssymb}
\usepackage{amsfonts}
\usepackage{amssymb}
\usepackage{dcolumn}
\usepackage{bm}
\usepackage{bbm}
\usepackage{graphicx}
\usepackage{xcolor}
\usepackage{array}
\usepackage{subfigure}
\usepackage{hyperref}
\usepackage{wasysym}

\usepackage{ulem}

\newcommand{\be}{\begin{equation}}
\newcommand{\ee}{\end{equation}}
\newcommand{\ba}{\begin{eqnarray}}
\newcommand{\ea}{\end{eqnarray}}

\newcommand{\lag}{\langle}
\newcommand{\rag}{\rangle}

\newcommand{\gsim}{\mathrel{\hbox{\rlap{\lower.55ex \hbox {$\sim$}}
                   \kern-.3em \raise.4ex \hbox{$>$}}}}
\newcommand{\lsim}{\mathrel{\hbox{\rlap{\lower.55ex \hbox {$\sim$}}
                   \kern-.3em \raise.4ex \hbox{$<$}}}}

\hypersetup{colorlinks=true,
            breaklinks=true,
            pdfstartview=Fit,
            linkcolor=blue,
            citecolor=blue,
            urlcolor=blue}

\begin{document}
\title{Probing new physics with multi-vacua quantum tunnelings beyond standard model through gravitational waves}

\author{Zihan Zhou}
\email{ustczzh@mail.ustc.edu.cn}
\affiliation{Department of Astronomy, School of Physical Sciences, University of Science and Technology of China, Hefei, Anhui 230026, China}
\affiliation{CAS Key Laboratory for Researches in Galaxies and Cosmology, University of Science and Technology of China, Hefei, Anhui 230026, China}
\affiliation{School of Astronomy and Space Science, University of Science and Technology of China, Hefei, Anhui 230026, China}

\author{Jun Yan}
\email{junyan96@mail.ustc.edu.cn}
\affiliation{Department of Astronomy, School of Physical Sciences, University of Science and Technology of China, Hefei, Anhui 230026, China}
\affiliation{CAS Key Laboratory for Researches in Galaxies and Cosmology, University of Science and Technology of China, Hefei, Anhui 230026, China}
\affiliation{School of Astronomy and Space Science, University of Science and Technology of China, Hefei, Anhui 230026, China}

\author{Andrea Addazi}
\email{andrea.addazi@lngs.infn.it}
\affiliation{Center for Theoretical Physics, College of Physics Science and Technology, Sichuan University, 610065 Chengdu, China}
\affiliation{INFN sezione Roma {\it Tor Vergata}, I-00133 Rome, Italy, EU}

\author{Yi-Fu Cai}
\email{yifucai@ustc.edu.cn}
\affiliation{Department of Astronomy, School of Physical Sciences, University of Science and Technology of China, Hefei, Anhui 230026, China}
\affiliation{CAS Key Laboratory for Researches in Galaxies and Cosmology, University of Science and Technology of China, Hefei, Anhui 230026, China}
\affiliation{School of Astronomy and Space Science, University of Science and Technology of China, Hefei, Anhui 230026, China}

\author{Antonino Marciano}
\email{marciano@fudan.edu.cn}
\affiliation{Department of Physics \& Center for Field Theory and Particle Physics, Fudan University, Shanghai 200433, China}
\affiliation{Laboratori Nazionali di Frascati INFN, Frascati (Rome), Italy, EU}

\author{Roman Pasechnik}
\email{Roman.Pasechnik@thep.lu.se}
\affiliation{Department of Astronomy and Theoretical Physics, Lund University, 221 00 Lund, Sweden, EU}

\begin{abstract}
\noindent
We report on a novel phenomenon of particle cosmology, which features specific cosmological phase transitions via quantum tunnelings through multiple vacua. The latter is inspired by the axiverse ideas, and enables to probe the associated new physics models through a potential observation of specific patterns in the stochastic gravitational waves background. Multiple vacua may induce the nucleation of co-existing bubbles over the phase transition epoch, hence enhancing the overall process of bubbles' nucleation. Our detailed analysis of semi-analytical and numerical solutions to the bounce equations of the path integral in three vacua case has enabled us to determine the existence of three instanton solutions. This new mechanism of cosmological phase transitions clearly predicts a possibly sizeable new source of gravitational waves, with its energy spectrum being featured with particular patterns, which could be probed by the future gravitational wave interferometers.
\end{abstract}




\maketitle

{\it Introduction.} --
Cosmological phase transitions (PTs) offer an inspiring possibility to probe physics beyond the Standard Model (SM). If first-order PTs took place at early cosmological times, a gravitational waves (GWs) spectrum can be induced, with crucial observational consequences for current and future GW experiments \cite{Witten:1984rs, Hogan:1984hx, Hogan:1986qda, Turner:1990rc}. Such a scenario is traditionally considered in the hot cosmological plasma characterized by a scalar-field effective potential accounting for both the loop and thermal corrections. In such a thermal system, the quantum tunneling is either ignored or only considered as happening between two vacua at a typical time scale of a given PT \cite{Quiros:1999jp}. However, in order to realize strong enough first-order PTs, several extended models of particle physics were considered that involve more than two vacua in the effective potential due to the presence of more degrees of
freedom. These are models with extra scalar singlets \cite{Anderson:1991zb, Espinosa:1993bs, Espinosa:2007qk, Profumo:2007wc, Espinosa:2011ax, Huang:2015bta} and doublets \cite{Cline:1996mga, Dorsch:2013wja, Basler:2016obg, Dorsch:2017nza, Basler:2017uxn}, as well as supersymmetric models \cite{Apreda:2001us, Huber:2007vva, Huber:2015znp}, and other SM extensions \cite{Cai:2016hqj, Hashino:2018wee}.

Nonetheless, new physics beyond the SM may arise for example by invoking the axion, originally postulated in Refs.~\cite{Peccei:1977hh, Weinberg:1977ma, Wilczek:1977pj} to address the strong CP problem in quantum chromodynamics (QCD), as well as the axion-like particles or the string axiverse scenario \cite{Arvanitaki:2009fg, Arvanitaki:2010sy} realized in plenty of UV theories. The axion was recently revitalized in the cosmological {\it relaxation} model, to dynamically address the electroweak (EW) hierarchy problem \cite{Graham:2015cka, Espinosa:2015eda, Gupta:2015uea}, which also naturally yields multiple vacua for a single scalar field. Inspired by this innovative phenomenology, we propose in this Letter to study quantum tunneling transitions, within the case of non-degenerate multiple vacua in a simple model with a single axion-inspired scalar field. Besides, we illustrate the possibility of probing new physics scenarios of this type by analysing the signals of the primordial GWs spectra generated by such transitions.

Instanton methods, initially developed in Refs.~\cite{Coleman:1977py, Callan:1977pt, Coleman:1980aw} to investigate quantum tunnelings in a gravitational environment, are nowadays widely exploited in the community. Even the functional Schr\"{o}dinger equation, supplied with the WKB approximation, was established to gain further insights into fields' potentials endowed with multi-vacua \cite{Sarangi:2007jb, Copeland:2007qf,Tye:2009rb}. The path integral over the quantum field configurations is addressed in terms of {\it the most probable escape paths} (MPEP), i.e. the instanton solutions given by the stationary phase approximation which dominates in the classically forbidden region. In this Letter, we perform a complete analysis of quantum tunnelings for the non-degenerate multi-vacua case, which preserves the major characteristics of the cosmological relaxation, and can yield well-behaved approximate solutions to the so-called bounce equations of the path integral. Around the reheating epoch\footnote{Several scalar fields can be accounted for giving rise to cosmological PTs, not only the inflaton. For instance, in the early Universe the curvaton may be taken into account \cite{Lyth:2001nq, Lyth:2002my}, requiring accordingly a curvaton reheating mechanism \cite{Feng:2002nb, Liddle:2003zw}.}, multiple types of bubbles would run away in a cosmic medium, and generate GWs that are expected to be examined in various observational windows, including GW astronomy and cosmic microwave background (CMB) signals \cite{Gordon:2002gv}.

{\it Multi-vacua quantum tunnelings.} --
In the string axiverse scenario, the axion-inspired scalar field $\phi$ evolves into the classically stable but quantum metastable regions where quantum fluctuations would become dominant in its subsequent evolution. For simplicity, we illustrate the three vacua case, assuming the decay through the four-vacua configuration in a single transition to be exponentially suppressed. Capturing the essence for multi-vacua quantum tunnelings, we consider the simplest scalar potentials as
\begin{equation} \label{curV}
\begin{split}
 \bar V(\bar\phi)=\left\{
 \begin{aligned}
 & \frac{1}{4} (\bar\phi+\sqrt{2}\bar m_{\phi})^2 \bar\phi^2 - b_1 (\bar\phi+\sqrt{2}\bar m_{\phi}) ~\Big|_{\bar\phi<0} , \\
 & \frac{1}{4} (\bar\phi-\sqrt{2}\bar m_{\phi})^2 \bar\phi^2 - (\sqrt{2}b_1\bar m_{\phi} + b_2\bar \phi) ~\Big|_{\bar\phi\geq0} . \\
 \end{aligned}
 \right.
\end{split}
\end{equation}
where bar denotes the quantities rescaled by a certain energy scale $\Lambda$. In Eq.~\eqref{curV}, three parameters were introduced, $\bar m _{\phi}$ roughly denoting the effective mass of the $\bar \phi$ field (in dimensionless units) at its false and true vacua, while $b_1$ and $b_2$ realize the energy differences among the three vacua. In our analysis, we impose matching conditions, avoiding any discontinuities in the model.

Since this axion-inspired scalar field $\phi$ was practically decoupled from any other SM fermions and bosons, while having suppressed its scalar self-interactions, 
the bubble nucleation temperature $T_n$ of early Universe would just provide the average kinetic energy available for the scalar field, without any significant modifications 
to the potential barrier shape. On the other hand, it is not always the case that we could find the temperature satisfying the thermal transition condition 
$S_3(T_n)/T_n\simeq O(100)$. Then, the dominant contribution to the PTs will be realized by quantum tunnelings with $T_n \approx 0$. The small thermal fluctuations 
here slightly reduce the energy difference in Eq.~\eqref{curV}, thus leading to a secondary effect compared to the quantum tunnelings. This model typically corresponds 
to a moment right after inflation, but before reheating, featuring small thermal corrections.
\begin{figure}[h!]
	\centering
	\includegraphics[width=0.45\textwidth]{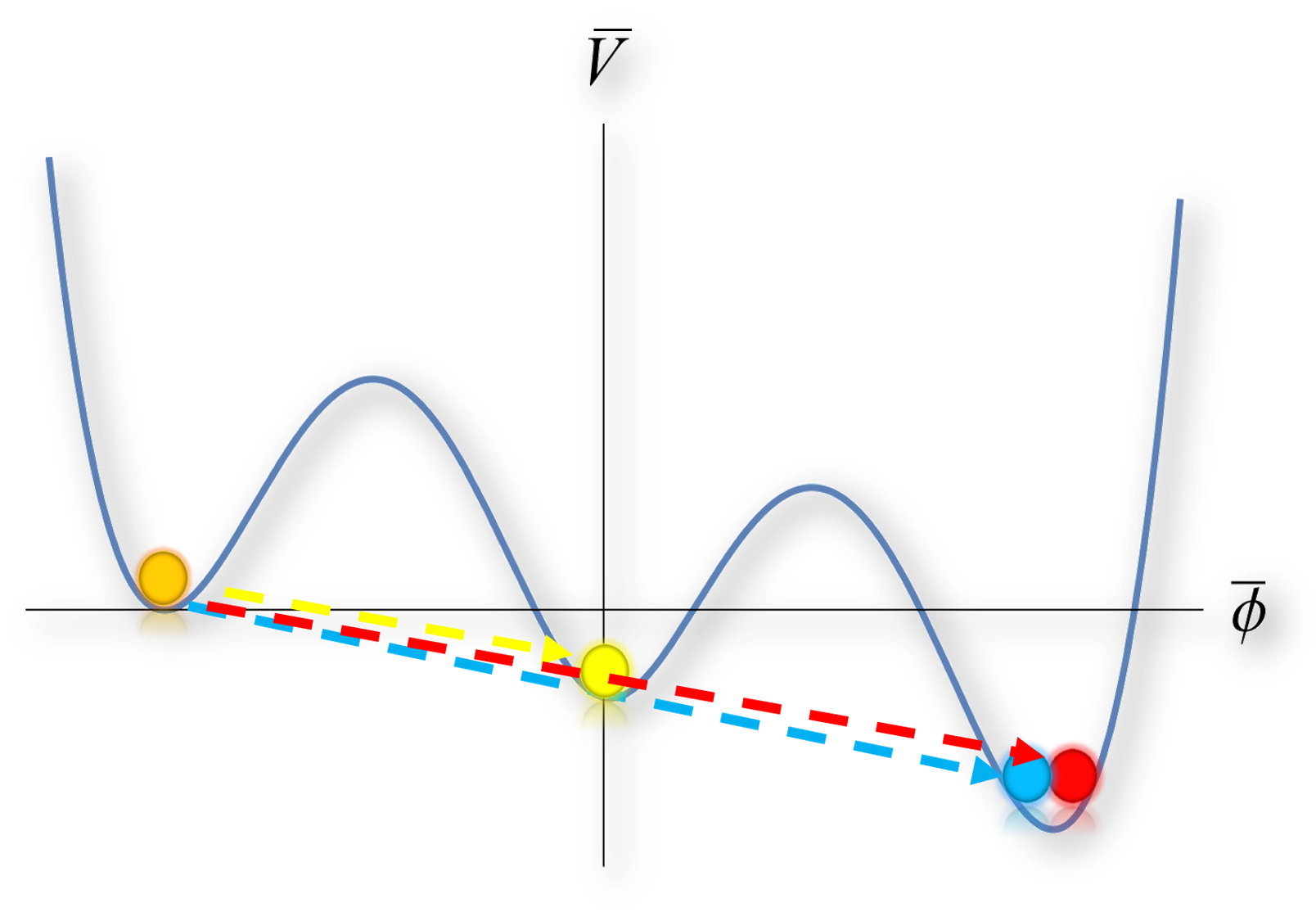}
	\includegraphics[width=0.45\textwidth]{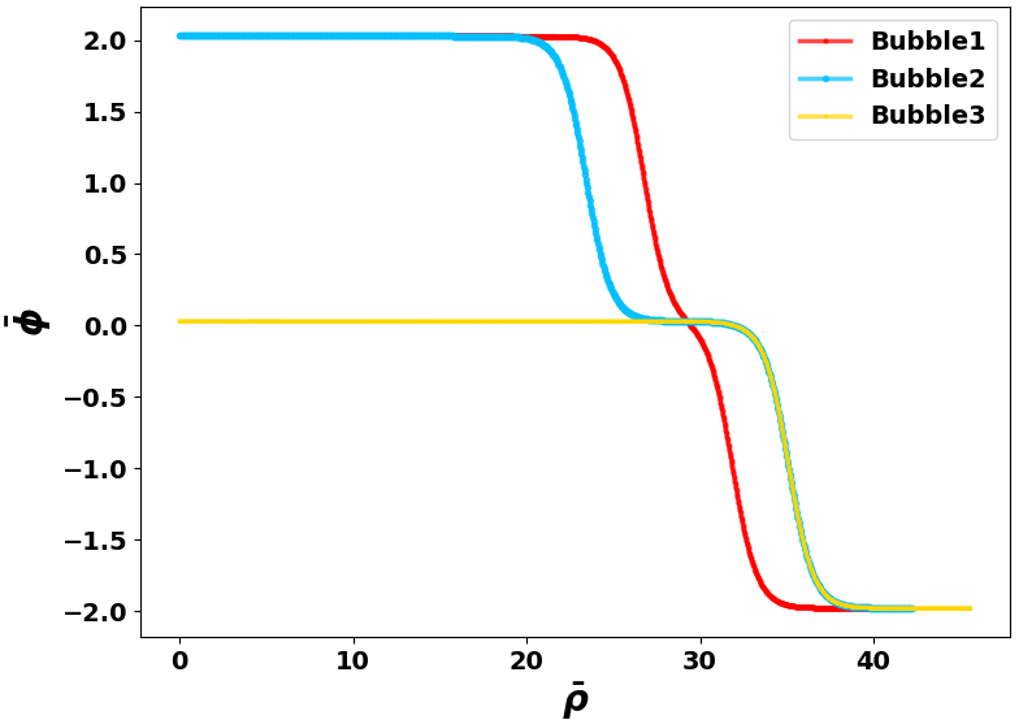}
	\caption{
		Upper: The potential $\bar V(\bar\phi)$ captured in Eq.~\eqref{curV}. The colourful dotted arrows depict the three MPEPs in our model. 
		Lower: The three bubble profiles with parameter choice: $\bar m_{\phi}=\sqrt{2}$, $b_1=0.04$, $b_2=0.06$.}
	\label{fig1:Potential}
\end{figure}

The profile of the potential is sketched in the upper panel of Fig.~\ref{fig1:Potential} with $\phi_F$, $\phi_M$ and $\phi_T$ to be the false, middle and true vacua, respectively. 
Ignoring thermal perturbations, quantum tunnelings originate from the homogeneous Universe in a false vacuum, and eventually terminate at the homogeneous Universe 
in a true vacuum.

The semiclassical equation of motion can be solved both numerically and analytically. With the functional Schr\"{o}dinger equation for
$\hat{H} \!=\! \int d^3 \mathbf{x} \big( - \frac{1}{2}(\frac{\delta}{\delta\phi(\mathbf{x})})^2 + \frac{1}{2}({\nabla}\phi)^2 + V(\phi)\big)$, namely
$\hat{H} \Psi(\phi(\mathbf{x})) \!=\! E \Psi(\phi(\mathbf{x}))$,
and the WKB approximation $\Psi(\phi) \!=\! A e^{i S(\phi)}$, with $S(\phi) \!=\! S_{(0)}(\phi) \!+\! S_{(1)}(\phi) \!+\! \cdots$, one may derive the MPEP 
semi-classical bounce equation
\begin{align}\label{bounce}
\begin{split}
 \frac{\partial^2\phi(\mathbf{x},\tau)}{\partial\tau^2} + \nabla^2\phi(\mathbf{x},\tau) -\frac{\partial V(\phi(\mathbf{x},\tau))}{\partial\phi} &=0 ~\Big|_{\tau < 0} ~,  \\
 \frac{\partial^2\phi(\mathbf{x},\tau)}{\partial \tau^2} - \nabla^2\phi(\mathbf{x},\tau) + \frac{\partial V(\phi(\mathbf{x},\tau))}{\partial\phi} &=0 ~\Big|_{\tau \geq 0} ~,
\end{split}
\end{align}
where $\tau$ is the parameter of MPEP, ranging from $-\infty$ to $+\infty$, with the critical point, separating the classically forbidden region ($\tau<0$) and the classically allowed region ($\tau \geq 0$), being fixed exactly at $\tau=0$. For cosmological PTs, the former equation can be recognized to describe the bubble nucleation processes, while the latter one drives the bubbles' evolution, during which the energy-momentum tensor of the field can evolve and generate GWs.

Following Ref.~\cite{Coleman:1977th}, the MPEP solutions to the first bounce equation ought to obey the $O(4)$ invariance and satisfy $\frac{\partial \phi}{\partial \tau}\big|_{\tau=0}=0$ 
as well as $\phi(\tau\rightarrow-\infty)=\phi_{F}$. The analytical solutions can be derived through the variational method:
\begin{equation}\label{trial}
\begin{split}
 \bar\phi(\bar\rho) = & \frac{1}{2} \bar\phi_{F} \tanh(\frac{\bar m_\phi}{2}(\bar\rho-\bar R_1)) + \frac{1}{2} \bar\phi_{F} \\
 & - \frac{1}{2} \bar\phi_{T} \tanh(\frac{\bar m_{\phi}}{2}(\bar\rho-\bar R_2)) + \frac{1}{2}\bar\phi_{T} ~,
\end{split}
\end{equation}
with $\bar\rho=\sqrt{{\bar{\mathbf{x}}}^2+\bar\tau^2}$. Here, $\bar R_1$, $\bar R_2$ are the parameters representing the bubble radii. Then, the Euclidean action can be expressed as
\begin{equation}\label{SE}
 S_E[\bar\phi] = 2\pi^2\int_{0}^{+\infty}{\bar \rho^3} \Big[ \frac{1}{2} \Big(\frac{d\bar\phi}{d\bar\rho}\Big)^2 + \bar V(\bar\phi) \Big] d\bar\rho ~.
\end{equation}
\begin{figure}[ht]
	\centering
	\includegraphics[width=0.5\textwidth]{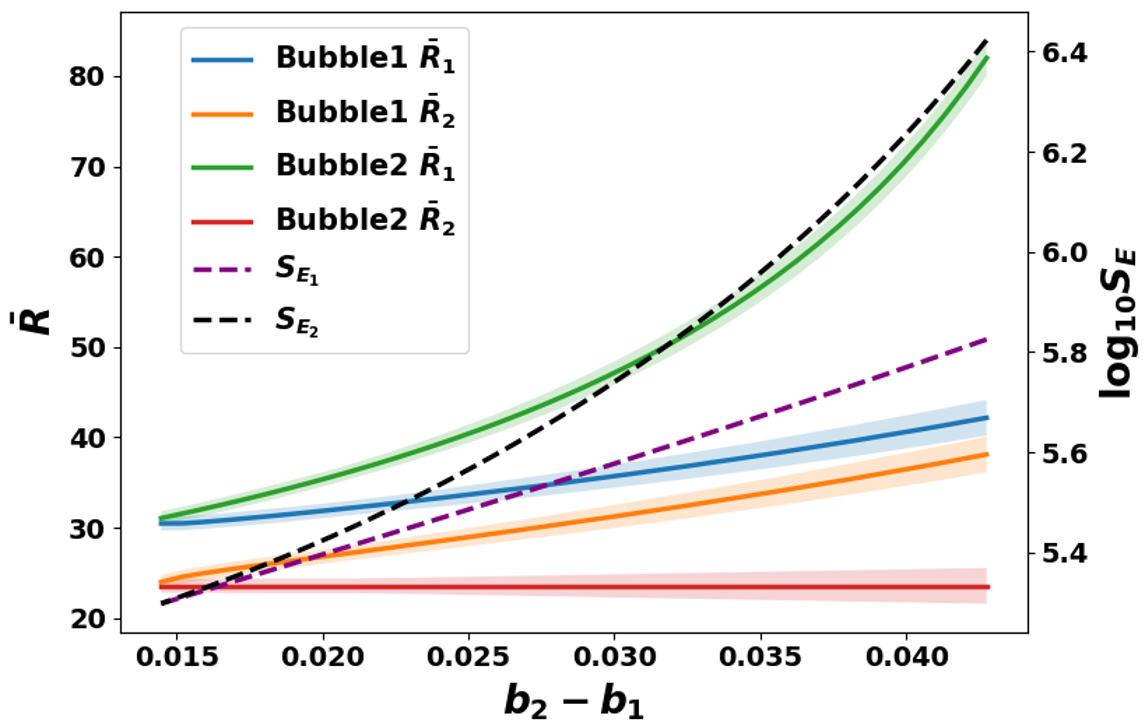}
	\caption{
		`Bubble1' and `Bubble2' dependence of the parameter $b_2-b_1$ after fixing $\bar m_{\phi}=\sqrt{2}$ and $b_2=0.06$. The bubble radius parameters $\bar R_1$ and $\bar R_2$ for the two solutions derived from variational methods are plotted in solid lines, with errors with respect to the exact solutions shown in the colored shadow area. Dashed lines represent the values of $S_E$ calculated in Eq.~\eqref{SE}.}
	\label{fig2:Bubble}
\end{figure}

Numerical computation can be derived via the Runge-Kutta algorithm for nonlinear ODE based on the package of CosmoTransitions \cite{Wainwright:2011kj}. The process involving three vacua provides us with a novel picture of tunneling transitions. In the previous literature, authors considered only one solution to the bounce equation, i.e. providing only one instanton solution. Our model can now lead to three instantons at most. The first obvious one is that if we require $\phi\big|_{\tau=0}=\phi_{M}$, we can get a trivial solution which is `Bubble3' in the bottom panel in Fig.~\ref{fig1:Potential}. This solution always exists as long as we have at least two non-degenerate vacua, while the other two solutions depend on the potential structure, especially the energy difference between ($\phi_{F}$, $\phi_{M}$) and ($\phi_{M}$, $\phi_{T}$) configurations.

All numerical solutions to the bounce equation are shown in the bottom panel of Fig.~\ref{fig1:Potential}. The three tunneling processes can occur simultaneously in the Universe, at an estimated rate per volume $\Gamma \sim e^{-S_{E}}$. In our case, `Bubble1' and `Bubble2' solutions driven by quantum effects add two new instantons, which cannot appear in the classical case. Since the thickness of bubble walls is small enough $\delta l/l \ll1$, the solutions can be further simplified in {\it thin-wall approximation} with the $\tanh$ function, in analogy with the solution to the quartic potential with degenerate vacua. Additionally, we see that the two radii of non-trivial solution `Bubble1' are nearly the same, which indicates that the false vacuum, under certain condition, could directly tunnel to the true vacuum without being significantly influenced by the mid-vacuum. In the `Bubble2' case, the solution can be simply derived by summing over the two instantons from the two vacua tunneling between ($\phi_{F}$, $\phi_{M}$) and ($\phi_{M}$, $\phi_{T}$), but finished in a single step transition in the three vacua case. Although we can also get more complicated cases like nested and reoccurring \cite{Vieu:2018zze} bubble profiles by only preserving O(3) symmetry and solving both of the two equations in Eq.~\eqref{bounce} simultaneously, these solutions are not from pure quantum effects, inconsistent with our main consideration. Note that the Universe was extremely empty after inflation and before reheating. In this epoch, the plasma effect is negligible, and the bubbles expand and collide with each other to reach thermal equilibrium. Since both `Bubble1' and `Bubble2' trigger the tunnelings towards $\phi_{T}$, the whole Universe shall reach the final state represented by the true vacuum, eventually.

Astonishingly, when the energy difference between $(\phi_F, \phi_{M})$ and $(\phi_{M},\phi_{T})$ captured by $b_2-b_1$ becomes small enough, `Bubble1' and `Bubble2' will vanish at the same time, leaving `Bubble3' the only solution. We name this novel phenomenon as ``{\it Two Step Tunneling}'' (TST). This is the first time that someone realized TST in cosmological PTs within the context of single field models\footnote{We refer to \cite{Wan:2018udw} for the realization with a two-field model and \cite{Vieu:2018zze, Addazi:2019dqt} for more exotic situations.}. All the related results are summarized in Fig.~\ref{fig2:Bubble}. Phenomenologically, the radius $R_2$ decreases faster than $R_1$ as $b_2-b_1$ decreases. When $b_2-b_1$ reaches certain critical point, the non-trivial MPEPs 'Bubble1' and 'Bubble2' vanish simutaneously. Numerically, in the given example, we scan the $b_2-b_1$ parameter space from $0.014$ to $0.043$, below which 'Bubble1' and 'Bubble2' disappear. From theoretical point of view, the behaviour of the Euclidean action $S_E$ determines all the phenomena related to these two solutions. It is easy to find that the stationary solution 'Bubble1' corresponds to the saddle point of $S_E$, while 'Bubble2' corresponds to the maximum point. For small energy difference between ($\phi_F,\phi_M$) and $(\phi_M,\phi_T)$, the maximum point and the saddle point will coincide. Thus for smaller $b_2-b_1$, there will be no static point for any bubble radius. In this case, `Bubble3' is the only MPEP ensuring the quantum decay. The false vacuum $\phi_{F}$ would first tunnel to the intermediate vacuum $\phi_{M}$, and then experience a second tunneling to reach the true vacuum $\phi_{T}$.

Even though TST may happen in the currently considered simple model, it may not occur in a realistic background. For instance, in the very early Universe, the curvaton could decay into radiation, and hence raise the temperature of the Universe thus approaching the reheating regime. In this case, a finite temperature correction $\sim T^2\phi^2 $ should be taken into account in the effective potential. After its quantum decay into the intermediate vacuum $\phi_{M}$, along with the increase of the temperature, the true vacuum $\phi_{T}$ would become degenerate with $\phi_{M}$, and then even vanish. If one still requires TST to occur in the specific example of Fig.~\ref{fig2:Bubble}, the reheating temperature is expected to be constrained by $\bar T\lsim 0.8\sqrt{b_2/\bar m_{\phi}}$, so that $\phi_{M}$ and $\phi_{T}$ would not be degenerate. As in this Letter we focus on a preliminary analysis of quantum tunnelings via multiple vacua, we leave to forthcoming studies more detailed investigations.

{\it {\rm GW} signals.} --
After the detailed analysis for the phase transition process, we give a preliminary consideration for its possible observational window which may interest future experiments. In the very early Universe background, the energy stored in the walls of vacuum bubbles during the expansion can effectively influence the curvature of spacetime and thus may be transfered to the gravitational radiation. In general, there exist three major sources for GWs from cosmological PTs \cite{Caprini:2015zlo, Caprini:2019egz}, which respectively are collisions of vacuum bubbles \cite{Kamionkowski:1993fg, Huber:2008hg}, sound waves \cite{Hindmarsh:2013xza, Hindmarsh:2017gnf} due to bubbles' expansions inside the plasma, and MHD turbulence \cite{Kosowsky:2001xp, Caprini:2009yp} after collisions. Concerning the PTs dynamics of our axion-inspired model in the vacuum-dominated epoch, which is not significantly affected by the thermal corrections as we have discussed, we naturally explore the bubble dynamics in the {\it run-away regime}, where it is well known that contributions from MHD turbulence and sound waves are negligible compared to collisional contributions \cite{Caprini:2015zlo, Caprini:2019egz}. As for the GWs production, it is sensitive to the temperature at reheating due to those different types of bubble dynamics and cosmological background. We assume that the GWs are produced in a thermal bath at temperature $T_{*}$ which approximately equals the reheating temperature $T_{*} \sim T_{\rm reh}$.

Recall that, the observational constraint upon reheating temperature is pretty loose, namely, the Big Bang Nucleosynthesis (BBN) yields a lower bound $T_{\rm reh} > 1~{\rm MeV}$ \cite{Kawasaki:2017bqm, Mielczarek:2010ag}. In this case, the GW intensity $\Omega_{\phi}$ and the related peak frequency $f_{\phi}$ caused by $\phi$ can be approximated as follows,
\begin{align}
\label{GW_Omega}
 \Omega_{\phi} h^2 \simeq & 1.67 \times 10^{-5} \Big(\frac{1}{\tilde\beta}\Big)^2 \Big(\frac{\kappa_{\phi}\alpha}{1+\alpha}\Big)^2 \Big(\frac{100}{g_*}\Big)^{\frac{1}{3}} \nonumber\\
 & \times \Big(\frac{0.11v_w^3}{0.42+v_w^2}\Big) \frac{3.8(f/f_{\phi})^{2.8}}{1+2.8(f/f_{\phi})^{3.8}} ~, \\
\label{GW_F}
 f_{\phi} \simeq & 1.65 \times 10^{-5} \mathrm{Hz} \Big(\frac{0.62}{1.8-0.1v_w+v_w^2}\Big) \tilde\beta \nonumber \\
 & \times \Big(\frac{T_*}{100\ \mathrm{GeV}}\Big) \Big(\frac{g_*}{100}\Big)^{\frac{1}{6}} ~,
\end{align}
with the two key parameters $\alpha$ and $\tilde\beta$. The former one, $\alpha$, is defined  at the PT temperature $T_n$ by $\alpha\! \equiv \!{\epsilon(T_n)}/{\rho_{\rm rad}(T_n)}$, and introduces the ratio between the energy difference among two vacua and the thermal energy density of the plasma $\rho_{\rm rad}(T_n) \propto T_n^4$. Considering the reheating background, the equation of state is $w=-1/3$ which corresponds to $H_n=1/t_n$. Then the rescaled bubble nucleation rate parameter $\tilde\beta$ can be captured by $\tilde\beta=\beta/H_n\simeq S_E$. Finally, $\kappa_{\phi}$ characterizes the fraction of the latent heat for the energy transfer.

\begin{figure}[ht]
\centering
\includegraphics[width=0.45\textwidth]{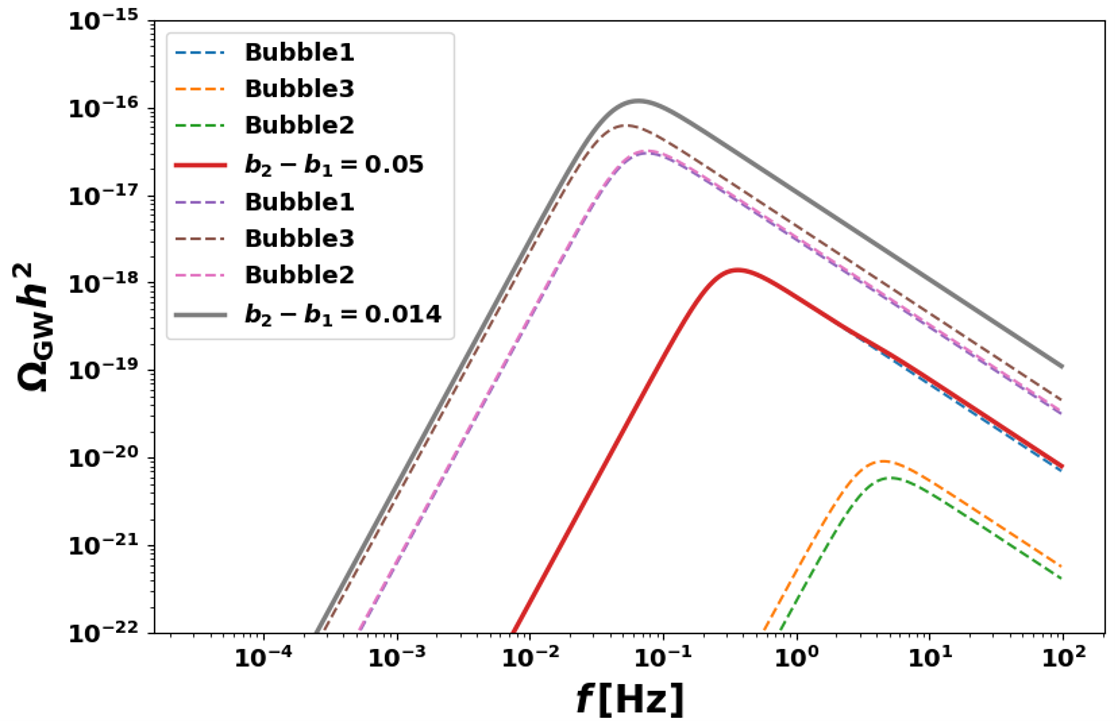}
\includegraphics[width=0.45\textwidth]{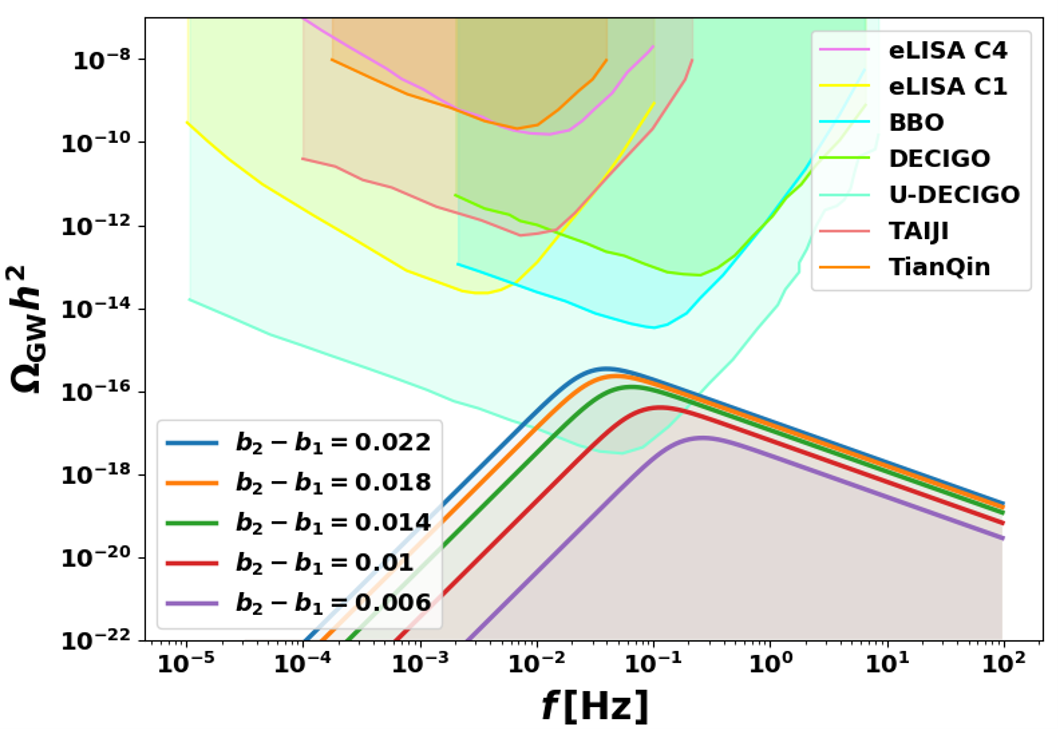}
\caption{
Upper: Contributions to GW spectrum from three bubbles in the given example are plotted in dashed lines; their summed total spectra are plotted in solid lines. 
Lower: GW intensity bands with different value choices of $b_2-b_1$ (which are $0.022$, $0.018$, $0.014$, $0.01$ and $0.006$ respectively). The solid lines are for the maximum intensities achieved at fixed $b_2-b_1$ while leaving $b_2$ and $\bar m_{\phi}$ free, and the top coloured regions correspond to the expected sensitivities of space-borne GW experiments including LISA \cite{Caprini:2015zlo}, BBO, DECIGO \cite{Moore:2014lga}, U-DECIGO \cite{Kudoh:2005as}, TAIJI \cite{Gong:2014mca} and TianQin \cite{Luo:2015ght}.}
\label{fig3:GW}
\end{figure}

PTs driven by $\phi$ in our model are due to quantum mechanical effects, implying $T_n \approx 0$ and thus $\alpha \rightarrow \infty$. In this limit, one gets $\Omega_{\rm GW} \approx \Omega_{\phi}$, with $\kappa_\phi \sim 1$ and $v_w \sim 1$. We present the results in Fig.~\ref{fig3:GW}. The total GW signals can be separated into three components, corresponding to the contributions of three types of vacuum bubbles. In the given example, for large energy difference between ($\sigma_{F}$, $\sigma_{M}$) and ($\sigma_{M}$, $\sigma_{T}$), i.e. $b_2-b_1 > 0.035$, the Euclidean actions for 'Bubble2' and 'Bubble3' are extremely enhanced and naturally lead to the GW energy spectrum generated by these two bubbles being severely suppressed. That results in the fact that the total spectrum is given by the `Bubble1' contribution. On the other hand, when $b_2-b_1$ is around $0.020$, the ratio $S_{E1}/S_{E2}$ is order of unity. The related tunneling rates will be of the same magnitude, giving rise to similar contributions from `Bubble1' and `Bubble2'. Generally, no matter which cases we are considering, the tiny differences between these three contributions can only influence the fine structure of the GW spectra, which could be examined thanks to the expected data from the GW interferometers in the future. Since $S_E$ significantly depends on $ b_2-b_1$, the amplitude of GW spectra can vary up to $O(10^2)$ for different model parameters in the considered example.
Although a similar shape of the primordial GW spectrum may also be realized within thermal PTs \cite{Ellis:2020awk}, we still emphasize that, the result predicted in our case is fundamentally different from the cosmological PTs of other types. For instance, in the standard case of thermal PTs there are extra contributions from sound waves and MHD turbulence, but all these become secondary in our case, which indicates that our scenario can be probed or falsified by the future high-sensitive GW interferometers.

{\it Conclusions.--}
In this Letter we put forward a novel mechanism to generate cosmological PTs via quantum tunneling transitions that may arise due to new physics described in terms of axiverse scenario with multiple vacua. Accounting for a specific solvable parameterization of the field potential, we made first semi-analytical and numerical analyses, providing an explicit solution involving three instanton configurations, and calculating the quantum decay rates. Our mechanism provides a platform for phenomenological investigations of the rich structure of quantum tunnelings, namely, an innovative realization of the TST phenomenon within the single field scenario. This process can lead to a spectrum of induced stochastic GWs, of which the fine structure is uniquely predicted. Due to the fact that its origin is different from that arisen from sound waves and MHD turbulence, this newly proposed GW source is observationally distinguishable in the future GW astronomy.

We end by discussing several implications of the novel mechanism that could inspire forthcoming studies. From theoretical perspective, our study illustrates that new physics beyond SM could be accessible through cosmological PTs if multiple vacua are allowed. This may be also related to the SM hierarchy problem, through embedding into the relaxation model. Phenomenologically, the fruitful phenomena in such transitions may lead to further considerations in stochastic GWs and PBH formations \cite{Khlopov:2008qy, Deng:2017uwc}. Also, we have neglected tunnelings along more consecutive vacua, but this theoretical possibility deserves further investigations, as a pathway to get better understanding of the new physics related to axion(-like) particles. Furthermore, the physical picture of quantum tunnelings can be also related to the inhomogeneous initial conditions that arise because of primordial perturbations, which may result in resonant tunnelings with higher decay rates. Although it may be challenging to test the heuristic example we are focused on within this Letter, with the resolutions of current GW experiments, our study can either be extended to several theoretical scenarios, or provide detection targets for the next generation of GW instruments.\\

{\it Acknowledgments.--}
We are grateful to Peter Graham, Misao Sasaki, Maxim Khlopov, Marek Lewicki, and Hao Yang for valuable comments.
YFC is supported in part by the NSFC (Nos. 11722327, 11653002, 11961131007, 11421303), by the CAST-YESS (2016QNRC001), by the National Youth 
Talents Program of China, and by the Fundamental Research Funds for Central Universities.
AM is supported in part by the NSFC (No. 11875113), by Shanghai Municipality (No. KBH1512299), and by Fudan University (No. JJH1512105).
R.P. was partially supported by the Swedish Research Council grant No.~2016-05996 and 
by the European Research Council (ERC) under the European Union's Horizon 2020 
research and innovation programme (grant agreement No 668679).
All numerics were operated on the computer clusters {\it LINDA} \& {\it JUDY} in the particle cosmology group at USTC.

\appendix
\section{SUPPLEMENTAL MATERIAL}

In order to give a detailed analysis of tunneling in quantum field theory, we need to specify the field theory version of WKB 
approximation \cite{Sarangi:2007jb, Copeland:2007qf,Tye:2009rb} analogous to the same approximation in quantum mechanics. 

The Hamiltonian of the scalar field $\phi(\textbf{x})$ in Schr\"{o}dinger picture reads
\begin{equation}
H=\int{d^3x\bigg(\frac{1}{2}\dot{\phi}^2+\frac{1}{2}({\nabla\phi})^2+V(\phi)\bigg)}~.
\end{equation}
To quantize the scalar field theory, we notice that the commutator of $\dot\phi$ and $\phi$ is given by 
$[\phi(\textbf{x}),\dot\phi(\textbf{x}')]=i\delta^3(\textbf{x}-\textbf{x}')$. Similar to quantum mechanics, 
we replace $\dot\phi$ by the operator $-i\delta/\delta\phi(\textbf{x})$ and consider the functional Schr\"{o}dinger equation
\begin{equation}
\hat{H}\Psi(\phi)=E\Psi(\phi)~,
\end{equation}
where the Hamiltonian operator can be written as 
\begin{equation}
\hat{H}=\int{d^3x\Bigg(
	-\frac{1}{2}
	\bigg(\frac{\delta}{\delta\phi(\textbf{x})}\bigg)^2
	+\frac{1}{2}({\nabla\phi})^2
	+V(\phi)\Bigg)}~.
\end{equation}
Here, $\Psi(\phi)$ should be interpreted as the wave function. It represents the probability of the occurrence of field operator $\phi$ in a configuration $\phi(\textbf{x})$. 
From here on, we talk about the scalar quantum field theory in configuration space. 

It proves useful to employ the WKB ansatz, $\Psi(\phi)=Ae^{iS(\phi)}$, $S(\phi)=S_{(0)}(\phi)+S_{(1)}(\phi)+\cdots$, where $A$ is a constant. Then we could derive
\begin{align}\label{WKBeq}
\begin{split}
\int{d^3\mathbf{x}\Bigg[
	\frac{1}{2}
	\bigg(\frac{\delta S_{(0)}(\phi)}{\delta\phi}\bigg)^2
	+\frac{1}{2}({\nabla\phi})^2
	+V(\phi)\Bigg]}=E~,\\
\int{d^3\mathbf{x}\Bigg[
	-i\frac{\delta^2S_{(0)}(\phi)}{\delta\phi^2}
	+2\frac{\delta S_{(0)}(\phi)}{\delta\phi}
	\frac{\delta S_{(1)}(\phi)}{\delta\phi}
	\Bigg]}=0~.
\end{split}
\end{align}
Similar to the formalism in quantum mechanics, we introduce the potential $U(\phi)$ with the definition
\begin{equation}
U(\phi)=\int{d^3x\bigg(\frac{1}{2}({\nabla\phi})^2+V(\phi)\bigg)}~.
\end{equation}
Then the classically forbidden region is defined as the region where $U(\phi)>E$, while the classically allowed region corresponds to $E>U(\phi)$. However, the difference from quantum mechanics is that Eq.~\eqref{WKBeq} is an infinite set of coupled nonlinear equations in the field configuration space, which is extremely hard to solve. To tackle this problem, the {\it the most probable escape paths} (MPEP) notion is introduced, which tells us that the tunneling probability is dominated by the contribution from a discrete set of paths in the configuration space. The nearby paths will only contribute through certain quantum corrections. In the current first analysis, we are only interested in the dominant contributions and ignore the quantum corrections for our purposes. By the definition of MPEP, any variation of $S_0$ perpendicular to this path should vanish, and along this path the variation of $S_0$ is nonvanishing. If we parameterize the path with $\tau$, we get 
\begin{align}
\frac{\delta S_{(0)}}{\delta\phi_\parallel}|_{\phi(\textbf{x},\tau)}&=C(\tau)\frac{\partial\phi}{\partial \tau}~,\\
\frac{\delta S_{(0)}}{\delta\phi_\perp}|_{\phi(\textbf{x},\tau)}&=0~,
\end{align}
where
\begin{equation}
C(\lambda)=
\frac{\partial S_{(0)}}{\partial\tau}
\Bigg(\int{d^3x
	\bigg[\frac{\partial\phi}{\partial\tau}\bigg]^2
	\Bigg)}^{-1}~.
\end{equation}
Then, after a lengthy calculation and refining the path parameter, we get the familiar Eq.~\eqref{bounce}. From another point of view, at the quantum level, after choosing the stationary phase approximation in the path integral formalism and doing the Wick rotation $\tau=it$, we could get the amplitude $\lag\phi_f|e^{-H\tau}|\phi_i\rag=Ae^{-S_E}$. Then, it is easy to find that the first equation in 
Eq.~\eqref{bounce} is exactly the stationary condition for the Euclidean action, and thus gives the dominant contribution to the path integral. 

However, unlike quantum mechanics, there is no resonant tunneling in the scalar field theory with the homogenous initial conditions, i.e. false vacuum for any space point. In the current framework, we can provide a proof of such no-go theorem. Similar to what we know in quantum mechanics, in order for resonant tunneling to occur, we need to have a middle classically allowed region, in which the classical Euler-Lagrangian equation holds. Now, generally, we have three widely accepted conditions for the resonant tunneling to occur:
\begin{align}
\bullet~&\mbox{A middle classically allowed region:}\nonumber\\
&E>U(\phi)\,;
\label{ClassAlllow}\\
\bullet~&\mbox{Energy conservation during quantum tunneling with} \nonumber\\&\mbox{homogenous initial condition:}\nonumber\\
&\phi(\tau\rightarrow-\infty)=\phi_F,\dot\phi(\tau\rightarrow-\infty)=0\,;
\label{Initial}\\
\bullet~&\mbox{the natural boundary conditions:}\nonumber\\
&\phi(\textbf{x}\rightarrow\infty)=\phi_F,\nabla\phi|_{\textbf{x}\rightarrow \infty}=0\,.
\label{Bound}
\end{align}

{\it Proof:} Given the arbitrariness in the choice of the potential-zero level, we set $V(\phi_F)=0$. Then we get the conserved energy $E=0$. 
Following Eq.~\eqref{ClassAlllow}, the two critical points $\tau_1,\tau_2$ at the conjection between the classically forbidden and allowed regions satisfy
\begin{equation}
\frac{\partial\phi(\tau,\textbf{x})}{\partial \tau}\bigg|_{\tau=\tau_1}=\frac{\partial\phi(\tau,\textbf{x})}{\partial \tau}\bigg|_{\tau=\tau_2}=0~.
\end{equation}
Then it is instructive to construct another integral $I(\textbf{x})$, with the definition
\begin{equation}
I(\textbf{x})=\int^{\tau_2}_{\tau_1}{d\tau\ \Bigg[V(\phi)
	-\frac{1}{2}\bigg(\frac{\partial\phi}{\partial \tau}\bigg)^2
	-\frac{1}{2}(\nabla\phi)^2\Bigg]}~.
\end{equation}
Making use of the Euler-Lagrangian equation, it is straightforward to find that $\nabla I=0$, which directly leads to 
$I(\textbf{x})=I(\infty)=0$. Combining this relation with $E=0$, we immediately see that 
\begin{equation}
\int d^3\textbf{x}{\int^{\tau_2}_{\tau_1}{d\tau\ 
		\Bigg[\bigg(\frac{\partial\phi}{\partial \tau}\bigg)^2
		+(\nabla\phi)^2
		\Bigg]}}=0~.
\end{equation}
It now follows that $\phi(\textbf{x},\tau_1\leq\tau\leq\tau_2)\equiv\phi_F$. This is a trivial solution to the Euler-Lagrangian equation meaning 
that the path has to be in the false vacuum which cannot describe any tunneling process.

However, in Ref.~\cite{Saffin:2008vi}, it is also pointed out that if we abandon the homogeneous initial condition constraint, we may realize the resonant 
tunneling in certain models. Studies of cosmological consequences of such a scenario would be an interesting project on its own for a future work.

\end{document}